# Liquid Metal Angiography for Mega Contrast X-ray Visualization of Vascular Network

**Qian Wang [1,4], Yang Yu [1,4], Keqin Pan [2], Jing Liu [1,3]\***

**1.** Department of Biomedical Engineering, School of Medicine,
Tsinghua University, Beijing, China

**2**. Hospital of Tsinghua University, Beijing, China.

**3.** Beijing Key Lab of CryoBiomedical Engineering and Key Lab of Cryogenics,
Technical Institute of Physics and Chemistry,
Chinese Academy of Sciences, Beijing, China

**4.** These authors contributed equally to this work.

**\*Address for correspondence:**
Dr. Jing Liu
Department of Biomedical Engineering,
School of Medicine,
Tsinghua University,
Beijing 100084, China
E-mail address: jliubme@tsinghua.edu.cn
Tel. +86-10-62794896
Fax: +86-10-82543767

**Abstract**

Visualizing the anatomical vessel networks plays a vital role in physiological or pathological investigations. However, identifying the fine structures of the smallest capillary vessels via conventional imaging ways remains a big challenge. Here, the room temperature liquid metal angiography was proposed for the first time to produce mega contrast X-ray images for multi-scale vasculature mapping. Gallium was used as the room temperature liquid metal contrast agent and perfused into the vessels of in vitro pig hearts and kidneys. We scanned the samples under X-ray and compared the angiograms with those obtained via conventional contrast agent--the iohexol. As quantitatively proved by the gray scale histograms, the contrast of the vessels to the surrounding tissues in the liquid metal angiograms is orders higher than that of the iohexol enhanced images. And the resolution of the angiograms has reached 100μm, which means the capillaries can be clearly distinguished in the liquid metal enhanced images. With tomography from the micro-CT, we also managed to reconstruct the 3-dementional structures of the kidney vessels. Tremendous clarity and efficiency of the method over existing approaches were experimentally demonstrated. It was disclosed that the usually invisible capillary networks now become distinctively clear in the gallium angiograms. This mechanism can be generalized and extended to a wide spectrum of 3-dimensional computational tomographic areas. It provides a soft tool for quickly reconstructing high resolution spatial channel networks for scientific researches or engineering applications where complicated and time consuming surgical procedures are no longer necessary.

## Introduction

Since William Harvey revealed the secret of blood circulation in the 17$^{th}$ century for the first time, physicians and scientists have discovered numerous facts and effects inside this cyclic system. Tremendous surgical efforts were ever made either in vivo or in vitro biological body, just wishing to see more about the distributions, functions and behaviors of various vascular networks. Generally, the circulation involves heart, lung, arteries, veins and capillaries, from which the blood can bring oxygen, nutrients to every organs and tissues and take away carbon-dioxide and all metabolic wastes. Meanwhile, for many clinical scenarios such as tumors and phlebeurysma, the abnormal growth of blood vessels may indicate progress of the disease [1, 2]. Further, tracing drugs and special molecules is also heavily dependent on understanding the vessels' anatomical structure. In fact, many clinical treatments, pathological evaluation as well as organ physiological interpretations are mostly carried out through the blood vessels into the road. Even for a typical biological or medical training, observing vasculature networks through surgical resections and microscope on selected animals is often a basic way for understanding the life sciences and technology. Overall, it is extremely critical to clarify the fine distribution and variation of the blood vessels in the physiological and pathological researches.

Angiography is a vascular reconstruction way that helps physicians diagnose and evaluate the physiological conditions related to blood vessels. It can be performed with many medical imaging methods, such as ultrasound (US), magnetic resonance imaging (MRI), computed tomography (CT) and X-rays [3-5]. US and MRI are mainly used for mapping large vessels. To improve the vascular image, many contrast agents that possess a different property to the surrounding tissues were often adopted. For CT and X-ray angiography, the efficacy of the contrast agent relies heavily on its density. Though both X-ray and CT can obtain rather high resolution 2D and 3D images, it remains a big challenge to clearly display the vessels, especially for those complex networks made up of tiny capillaries. So far, most of the existing contrast agents are made of solutions whose density is close to that of water. For such situations, the resembling property between the agent and the capillaries in tiny scale significantly reduces the possibility to distinguish them in the medical image.

Recently, tremendous efforts have been made to find new contrast agents which could significantly improve the vascular imaging quality. Among all the imaging methods ever tried, it is X-ray that was applied to angiography for the first time in 1896, just in the following year of Röntgen's discovery of the beam, in the study of a corpse's hand [6]. As an inexpensive and widely available imaging way, the basic idea for X-ray angiography is to modify the density in the vessels to distinguish them from the surrounding tissues. The most typical contrast media as already investigated generally include iodine and iodinated agents [7], nanoparticles [8-11], carbon dioxide [12] and so on. Such vascular-enhanced radiological angiography in fact becomes the foundation for CT and 3D vasculature reconstruction [13-15]. It is also a standard reference for evaluating other contrast-enhanced imaging methods [16-18]. Generally, high energy X-ray is a necessity for obtaining good contrast image. However, when raising the energy of the X-ray, the contrast efficacy of iodine or $CO_2$ would decrease.

As an alternative, the newly emerging nanoparticles were intensively investigated aiming to further improve the quality of the existing contrast agents. However, the improvement is still rather limited.

Here, instead of searching for and trying more complex chemicals, we proposed for the first time the liquid metal angiography to realize ever powerful vascular radiological imaging which offers mega contrast X-ray images as compared with conventional ways. Through infusing the room temperature liquid metal such as gallium into the coronary artery of the pig heart in vitro, we managed to significantly lighten the target vessels in several orders larger than before under the ordinary radiological imaging instrument. In this way, the capillaries used to be hardly detectable are now easily seen on the image with outstanding clarity. This method opens the mega contrast nonsurgical approach for characterizing the fine distribution and variation of the biological vasculature system. It also suggests possibility for a localized in vivo vascular-enhanced radiological imaging in the near future. The principle has generalized purpose and can be extended to a variety of 3-dimensional computational tomographic areas including many engineering applications other than the biomedical category alone. This work sets up a highly efficient soft tool for quickly reconstructing high resolution spatial channel network for scientific researches or engineering applications where complicated and time consuming resections are no longer needed.

## Materials and Methods

The present study has been approved by the Ethics Committee of Tsinghua University, Beijing, China under contract [SYXK (Jing) 2009-0022].

As the first trial in this area, the gallium metal with a purity of 99.999% was adopted as the contrast agent. There are several reasons for such choice. First, the melting point of gallium is 29.78 ℃, which is close to the room temperature and below the body temperature (37 ℃). Thus it is easy to transform the metal between its liquid and solid phase. What's more, the gallium endows a significant property of overcooling, which means the phase change does not happen immediately when the temperature drops below the melting point. It is thus rather beneficial for manipulating the metal flexibly. Second, the gallium is chemically very stable and does not react with water at around body temperature. Third, a series of former researches have proven that gallium is safe for human in many normal occasions. Most importantly, the density of gallium is much higher than that of the blood. Therefore, it has superior visibility under X-ray irradiation and will not infiltrate through or be washed away after injected into the target vessels.

Physiologically, heart is the engine of the circulatory system. Oxygen and nutrients to the heart is conveyed by the coronary artery, which runs throughout the myocardium with its complex capillary network supporting behind. Many diseases happen with the pathological changes in the heart vessels, and it is vital to sketch out the vessels' fine distribution. The kidney is another important organ with large amount of vessels, which plays the role of blood filter. Therefore, for such consideration, we choose the commercially available pig hearts and kidneys which were all within 2 days' fresh state after sacrifice as the test objects to investigate the basic imaging principle of the liquid metal based radiological angiography.

Repetitive imaging experiments were performed on each group of the above different organs using CT scanner (GE XR/A, Phillips Brilliance 6). And all the results disclosed the distinctively high contrast quality of the presented imaging method.

## Experimental Results

As is anticipated, the angiograms depicted in Fig. 1A and Fig. B display tremendous differences for the hearts perfused with liquid metal gallium and iohexol (35g(I)/100ml), respectively. Interestingly enough, the gallium produces superior contrast effect intuitively and enables one to see through a highly clear coronary network, while only several relatively large vessels could be observed in the heart perfused with iohexol as commonly used so far. To quantify such improvement, we calculated the gray scale curves at five different heights in both angiograms and compared them with each other. The curves of a1-a5 representing the gallium angiography case all show significantly steep peeks than those in b1-b5 of the iohexol enhanced image cases (Fig. 1C).

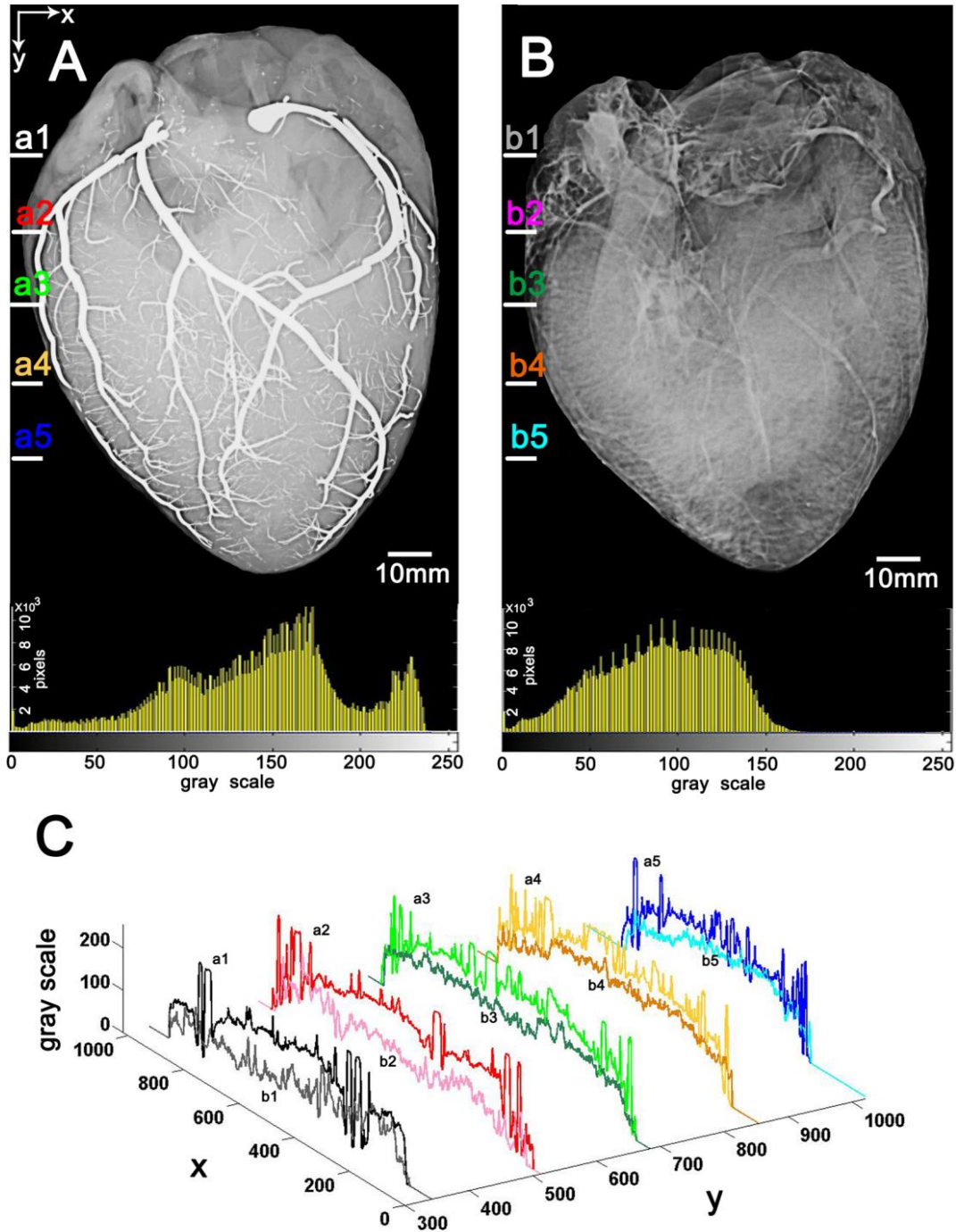

**Figure 1. Comparison of contrast effects between angiograms with liquid metal and conventional agent under X-ray irradiation.** (**A**) X-ray angiogram of heart filled with liquid metal gallium. (**B**) X-ray angiogram of heart filled with iohexol. Histogram of either image is placed at the bottom of the image A and B. (**C**) Plots of the gray scale along the horizontal lines at 5 different heights

labeled in A and B (For aesthetic, we did not draw the whole lines on images of A and B). The working parameters for the X-ray in both A and B are 55kV and 25mAs.

Figure 2 reveals the effects of X-ray irradiation intensity on the image quality. With the same exposure parameter (10mAs), the contrast between the vessels (gallium) and the other soft tissues increases along with the addition of the X-ray intensity (Fig. 2A, 80kV; Fig. 2B, 100kV; and Fig. 2C, 120kV). However, the increasing penetrability also makes some capillaries invisible, which can be seen in the sub-images in a1~a3 and b1~b3. The plots of the sub-images' gray scale histograms in Fig. 2D and Fig. 2E have given proofs in quantity. In both plots, there exist two peaks for the sub-images, which can be generally seen as one representing the enhanced vessels and the other representing the background tissues. The distance between the two peaks becomes longer when the X-ray gets stronger, indicating a higher contrast; while the height of the vessel peak decreases at the same time, which means the vessels become darker. Yet the illumination tends to saturate when the intensity is low, which will cause the loss of density information. Thus a higher intensity is more likely applicable for the angiographic X-ray computational tomography (CT) with gallium. Fig. 2F is the 3D reconstruction of the coronary arteries from the CT scan, whose parameters are set as 120kV and 100mAs.

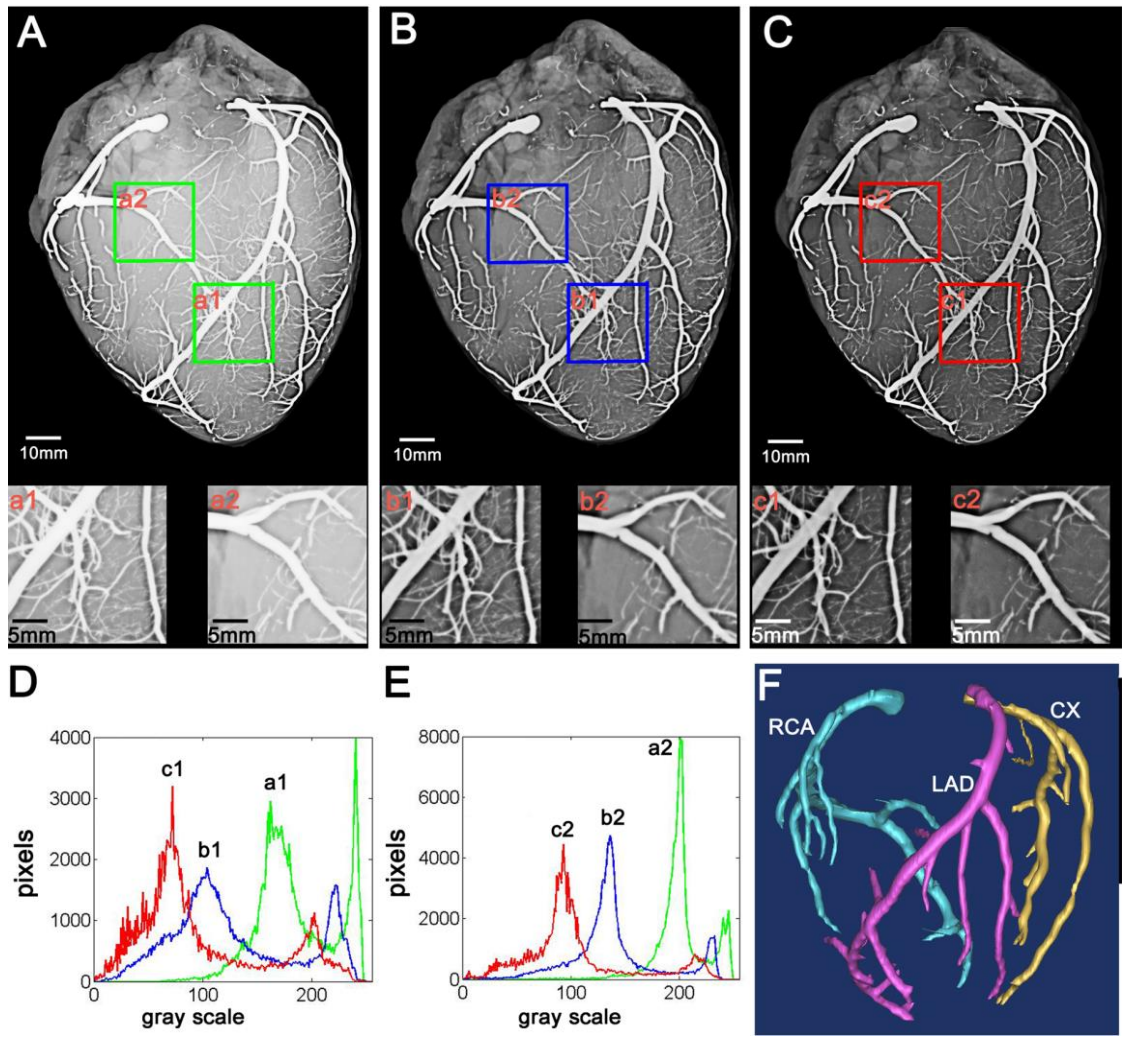

**Figure 2. Effect of X-ray irradiation intensity on image contrast of heart filled with liquid metal gallium.** (**A**) Image of the heart at 80kV. (**B**) Image of the heart at 100kV. (**C**) Image of the heart at 120kV. The exposure parameter in all 3 images is set as 10mAs identically. The square labeled with a1, a2, b1, b2, c1, c2 is zoomed in and presented below the image A, B, C, respectively. (**D**) Comparison of the histogram for three sub-images labeled with a1, b1, c1 in A, B and C. (**E**) Comparison of the histogram for three sub-images labeled with a2,b2,c2 in A,B and C. (**F**) Gallium enhanced coronary 3D vasculature of the heart, with the right coronary artery (RCA) in red, the left anterior

descending (LAD) in magenta and the circumflex (CX) in yellow. The parameters of the CT scan are 120kV and 55mAs.

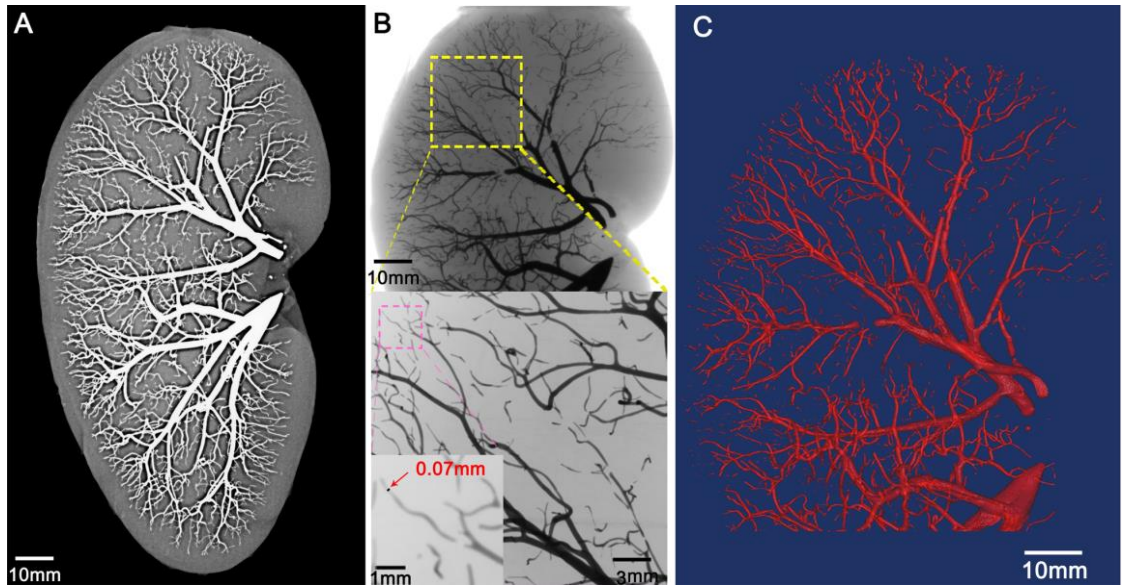

**Figure 3. Mega contrast vasculature of a pig kidney perfused with liquid metal gallium under X-ray irradiation.** (**A**) Angiogram of a whole pig kidney with its arterial network filled with gallium. The network of arterial vessels shows high contrast to the ambient tissues. The X-ray parameters for this angiogram are 45kV and 3.2mAs. (**B**) The 2D-view images scanned with micro-CT. The diameter of the thinnest vessels is down to 0.07mm. (**C**) Gallium enhanced kidney 3D vasculature.

In order to offer more evidences that the liquid metal angiography is highly suitable for mapping various organ vasculatures, we also managed to infuse the gallium into the renal artery of the pig kidney. The results are shown in Fig. 3A. The whole renal artery network is intact and the texture of the small vessels is rather clear. Therefore, segmentation of the renal artery becomes rather easy now by the new method. To obtain a clear image for these finer vascular, we further scanned this kidney with micro-CT (XM-Tracer-130, Institute of High Energy Physics, Chinese Academy of Sciences), and one can even see the thinnest tube in the images that has reached 100μm (Fig. 3B). In fact, given the resolution of the imaging apparatus is high enough, the liquid metal angiography allows to reconstruct further smaller size vessel (Fig. 3C) since such metal fluid could be injected to even nano tube.

## Discussion

Clearly, the introduction of the liquid metal angiography opened a brand new way for revolutionizing the current radiological imaging. The liquid phase of the metal warrants its easy flow into the capillaries, while the extremely high density of such special agent offers a mega contrast which is hard to achieve via existing approaches. Thus it is entirely feasible to map further smaller vessels using the present method on condition that the resolution of the imaging device allows for more precise scanning like the X-ray micro-CT [13, 19]. With such a mega contrast, it is rather easy and convenient to rebuild the fine vascular structure. The capacity of this imaging strategy is huge which can in fact find strong evidence from former nano fluidic researches. For example, a nanothermometer was once demonstrated with gallium filled in the carbon nanotube with a diameter of 75nm, yet the way to "read the scale" was recommended as the scanning electron microscope (SEM) [20].

Clearly, with the liquid metal angiograph for targeted nano-size vessel network, reading with a high-resolution X-ray will be feasible, too.

The extremely large X-ray attenuation coefficient of gallium is owing to its high density. Let $\mu_2$ and $\mu_1$ denote the X-ray attenuation coefficient of the contrast agent filled in the vessels and that of the surrounding tissues respectively, and $l$ denotes the diameter of a vessel. The contrast of this vessel in the whole image $C$ can then be expressed as

$$C=1-\exp[-(\mu_2-\mu_1)l] \qquad (1)$$

It tells that, under the same contrast condition, the larger the difference between $\mu_2$ and $\mu_1$, the clearer the vessel can be seen. Therefore, the high X-ray attenuation coefficient of the gallium makes it a mega contrast X-ray agent, which indicates that vessels can be more distinct in the image. However, the high attenuation property means that stronger X-ray irradiation intensity is needed. Otherwise, the X-ray cannot penetrate the whole objective. Generally, the attenuation is mainly caused by Compton scattering when the X-ray energy is high, then the attenuation factors of the contrast agent and the surrounding tissues will have small differences. Naturally, the corresponding image will produce poor contrast. However, for gallium angiography, this problem does not exist as reflected in Fig. 2.

Compared to all the other existing agents, the gallium is just a simple substance with no need of further processing or reaction. Thanks to its liquid phase around room temperature, the gallium is easy to flow into the capillaries, and its high density significantly raises the contrast of the vessels under the X-ray, which is obvious in the grey level histograms. On condition that the imaging apparatus allows, the tiniest channel that can be distinguished using gallium could even reach nano meter scale. With all the above merits, the metal flow performs far much better than those conventional contrast agents for visualizing the vessel networks, which suggests that the gallium could serve as an ideal contrast agent for in vitro vascular-enhanced radiological imaging.

The liquid metal has been provided as a highly convenient and useful tool for reconstructing the distribution of the fine vessel networks in tissues and organs, especially in computational tomography and 3D modeling, which is quite important for studying visual animal vasculature including human subjects in the near future. Previously, to reconstruct such anatomical vascular networks, tremendous efforts, times and costs are requested, as indicated by the well-known Vitual Human Program performed before throughout the world. Besides, since the liquid metal will overflow through the orifices, the infusion may help find out some unnoticeable wounds on or beneath the surface, which can be applied in the forensic detection. What's more, in vivo localized vascular-enhanced imaging is also promising. Since the gallium does not react with water and its high density makes it difficult to be washed away, a small amount of the metal can be infused to the target fine vessels in the living tissues and sucked out without residual.

Overall, the liquid metal angiography has demonstrated significant values in the field of vascular network visualization. The filling and freezing of the liquid metal would keep the shapes of the targets. Clearly, the basic idea of the present method is rather generalized. For example, the objects to be characterized can be extended to more other cavities spanning from animals' digestive tracts to plants' tube structures, even including insects' holes. Verification of such liquid metal based channel mapping may indicate more unexpected discoveries. Further, the unique properties of the liquid metal, such as a high conductivity of heat and electricity, also suggest that physical principles in these aspects can be considered following the perfusion to achieve different biomedical performances.

## Conclusion

In summary, the liquid metal angiograph as established in this study offers mega contrast quality for reconstructing the significantly enhanced radiological vascular imaging. With melting point around room temperature and pretty higher density over conventional image contrast agent, the liquid metal can be easily injected into the vascular system which would finally fill the finest capillaries and lighten them with superior clarity under the X-ray scan. This method is also highly applicable for computational tomography and 3-dimentional vasculature reconstruction, which will no longer need complex, expensive and time consuming surgical resections. It opens a highly powerful angiograph tool for physiological and pathological researches, which can also possibly be used for certain localized in vivo situations in the near future. In addition, owing to its generalized applicability, the present image reconstruction principle can even be extended to more other scientific or engineering areas where quickly reconstructing high resolution spatial channel networks is requested. As a fundamental discovery as well as an important step as made in the material enhanced angiography area, this study may also help refresh people's basic understandings on visualizing biological anatomy either in an organ or whole body and is expected to generate impact for future researches and practices. Lastly, from the device aspect, the present unconventional image enhancing method paved the way to exploit full potentials of X-ray that would significantly enhance its maximum performance for many scientific studies related to channel network reconstruction in the near future.

## References

1. Konerding MA, Fait E, Gaumann A (2001) 3D microvascular architecture of pre-cancerous lesions and invasive carcinomas of the colon. British Journal of Cancer 84: 1354-1362.

2. Carmeliet P, Jain KR (2011) Principles and mechanisms of vessel normalization for cancer and other angiogenic diseases. Nature Reviews Drug Discovery 10: 417-427.

3. Zhang Z, Wang H, Zhou Y, Wang J (2013) Computed tomographic angiography of anterior spinal artery in acute cervical spinal cord injury. Spinal Cord 51: 442-447.

4. Anderson CM, Saloner D, Lee RE, Griswold VJ, Shapeero LG, Rapp JH, Nagarkar S, Pan X, Gooding GA (1992) Assessment of carotid-artery stenosis by MR angiography-comparison with X-ray angiography and color-coded Doppler ultrasound. American Journal of Neuroradiology 13: 989-1003.

5. Knowles J (2003) Seeing is believing. Science 299: 2002-2003.

6. Darius J (1984) Radiography of the living brain. Nature 308: 225-225.

7. Hallouarda F, Antona N, Choquetb P, Constantinescob A, Vandammea T (2010) Iodinated blood pool contrast media for preclinical X-ray imaging applications - A review. Biomaterials 31: 6249-6268.

8. Rabin O, Perez JM, Grimm J, Wojtkiewicz G, Weissleder R (2006) An X-ray computed tomography imaging agent based on long-circulating bismuth sulphide nanoparticles. Nature Material 5: 118-122.

9. Liu H, Wang H, Guo R, Cao XY, Zhao JL, Luo Y, Shen MW, Zhang GX, Shi XY (2010) Size-controlled synthesis of dendrimer-stabilized silver nanoparticles for X-ray computed tomography imaging applications. Polymer Chemistry-Uk 1: 1677-1683.

10. Liu Y, Ai K, Liu J, Yuan Q, He Y, Lu L (2012) A High-performance ytterbium-based nanoparticulate contrast agent for in vivo X-ray computed tomography imaging. Angewandte Chemie International Edition 51: 1437-1442.

11. Hainfeld JF, Slatkin MD, Focella TM, Smilowitz HM (2006) Gold nanoparticles: A new X-ray contrast agent. British Journal of Radiology 79: 248-253.

12. Lundstrom U, Larsson DH, Burvall A, Scott L, U.K. Westermarkm, M. Wilhelm, M.A. Henriksson, H.M. Hertz (2012) X-ray phase-contrast $CO_2$ angiography for sub-10μm vessel imaging. Physics in Medicine and Biology 57: 7431-7441.

13. Ananda S, Marsden V, Vekemans K, Korkmaz E, Tsafnat N, Soon L, Jones A, Braet F (2006) The visualization of hepatic vasculature by X-ray micro-computed tomography. Journal of Electron Microscopy 55: 151-155.

14. Van den Wijngaard JP, Schwarz JC, Van Horssen P, Van Lier MG, Dobbe JG, Spaan JA, Siebes M (2013) 3D Imaging of vascular networks for biophysical modeling of perfusion distribution within the heart. Journal of Biomechanics 46: 229-239.

15. Jorgensen SM, Demirkaya O, Ritman EL (1998) Three-dimensional imaging of vasculature and parenchyma in intact rodent organs with X-ray micro-CT. American Journal of Physiology-Heart and Circulatory Physiology 275: H1103-H1114.

16. Yang Q, Li K, Liu X (2009) Contrast-enhanced whole-heart coronary magnetic resonance angiography at 3.0-T: a comparative study with X-ray angiography in a single center. Journal of the American College of Cardiology 54: 69-76.

17. Froeling V, Diekmann F, Renz DM, Fallenberg EM, Steffen IG, Diekmann S, Lawaczeck R, Schmitzberger FF (2013) Correlation of contrast agent kinetics between iodinated contrast-enhanced spectral tomosynthesis and gadolinium-enhanced MRI of breast lesions. European Radiology 23: 1528-1536.

18. Khilnani NM, Winchester PA, Prince MR, Vidan E, Trost DW, Bush HL, Watts R, Wang Y (2002) Peripheral vascular disease: Combined 3D bolus chase and dynamic 2D MR angiography compared with X-ray angiography for treatment planning. Radiology 224: 63-74.

19. Plouraboue F, Cloetens P, Fonta C (2004) X-ray high-resolution vascular network imaging. Journal of Microscopy-Oxford 215:139-148.

20. Gao YH, Bando Y (2002) Carbon nanothermometer containing gallium. Nature 415: 599-600.